\documentclass[12pt]{iopart}

\usepackage{dcolumn}
\usepackage{bm}         
\usepackage{amssymb}
\usepackage{graphicx}
\usepackage{color}

\newcommand{\gapprox}{{\scriptscriptstyle\stackrel{>}{\sim}}}
\newcommand{\lapprox}{{\scriptscriptstyle\stackrel{<}{\sim}}}

\newif\ifgraph

\graphtrue                   %

\begin{document}


\title[YBCO GBJs and low-noise SQUIDs patterned by FIB]{YBa$_2$Cu$_3$O$_7$ grain boundary junctions and low-noise superconducting quantum interference devices patterned by a focused ion beam down to 80\,nm linewidth}

\author{J.~Nagel$^1$, K.~B.~Konovalenko$^1$, M.~Kemmler$^1$, M.~Turad$^1$, R.~Werner$^1$, E.~Kleisz$^2$, S.~Menzel$^2$, R.~Klingeler$^2$, B.~B\"uchner$^2$, R.~Kleiner$^1$ and D.~Koelle$^1$ }

\address{$^1$ Physikalisches Institut -- Experimentalphysik II and Center for Collective Quantum Phenomena,
  Universit\"at T\"ubingen,
  Auf der Morgenstelle 14,
  D-72076 T\"ubingen, Germany}
\address{$2$ Leibniz-Institut f\"ur Festk\"orper- und Werkstoffforschung (IFW) Dresden,
  D-01171 Dresden, Germany}
\ead{koelle@uni-tuebingen.de}

\begin{abstract}
YBa$_2$Cu$_3$O$_7$ 24$^\circ$ (30$^\circ$) bicrystal grain boundary junctions (GBJs), shunted with 60\,nm (20\,nm) thick Au, were fabricated by focused ion beam milling with widths
$80\,{\rm nm} \le w \le 7.8\,\mu$m.
At 4.2\,K we find critical current densities $j_c$ in the $10^5\,{\rm A/cm^2}$ range 
(without a clear dependence on $w$) and an increase in resistance times junction area $\rho$ with an approximate scaling $\rho\propto w^{1/2}$.
For the narrowest GBJs $j_c\rho\approx 100\,\mu$V, which is promising for the realization of sensitive nanoSQUIDs for the detection of small spin systems.
We demonstrate that our fabrication process allows the realization of sensitive nanoscale dc SQUIDs; for a SQUID with $w\approx 100$\,nm wide GBJs we find an rms magnetic flux noise spectral density of $S_\Phi^{1/2}\approx 4\,\mu\Phi_0/{\rm Hz}^{1/2}$ in the white noise limit.
We also derive an expression for the spin sensitivity $S_\mu^{1/2}$, which depends on $S_\Phi^{1/2}$, on the location and orientation of the magnetic moment of a magnetic particle to be detected by the SQUID, and on the SQUID geometry.
For the not optimized SQUIDs presented here, we estimate $S_\mu^{1/2}=390\,\mu_B/\sqrt{\rm{Hz}}$, which could be further improved by at least an order of magnitude.
%
\end{abstract}



\maketitle

\section{Introduction}

There is a growing interest in developing sensitive miniaturized superconducting quantum interference devices (SQUIDs) for the investigation of small spin systems and scanning SQUID microscopy with sub-$\mu$m spatial resolution \cite{Wernsdorfer00,Gallop03,Sust09nanoSQUIDs}.
A main motivation are measurements on single nanomagnetic particles, and the ultimate goal is 
the direct detection of switching of a single electronic spin with various potential applications in spintronics, quantum computing and on biomolecules.
These applications require sub-$\mu$m Josephson junctions and SQUID loops, both for optimum inductive coupling to nanosized objects and for improving the SQUID sensitivity for the operation at switching fields of the magnetic particles up to the Tesla range
at temperature $T\approx 4\,$K and below (see e.g.~Refs.~\cite{Gallop03,Bouchiat09,Tilbrook09,Wernsdorfer09} and references therein).

Most frequently used techniques for creating superconducting sub-$\mu$m thin film structures are based on electron beam lithography, or on focused ion beam (FIB) patterning.
While the patterning of sub-$\mu$m SQUID loops poses no particular problems, the properties of sub-$\mu$m Josephson junctions embedded in the SQUID loop have to be carefully optimized in order to realize low-noise SQUIDs.
Here, an important figure of merit is the rms spectral density of flux noise $S_\Phi^{1/2}$, which for optimized SQUIDs based on the standard Nb/AlO$_x$/Nb technology for $\mu$m-sized junctions is in the range of $\mu\Phi_0/\sqrt{\rm Hz}$ ($\Phi_0$ is the magnetic flux quantum).
However, critical current densities $j_c>1\,{\rm kA/cm}^2$ are hard to achieve reliably with this technology, yielding too small critical currents $I_c=j_cA_J$ for junction areas $A_J$ in the $(100\,{\rm nm})^2$ range.
Therefore, recent research focused on Nb or Al thin film constriction type junctions (with widths $w\,\gapprox\,50\,$nm), sometimes shunted with a thin film normal metal layer (e.~g.~Au, or W) to ensure non-hysteretic current-voltage-characteristics (IVCs) \cite{Sust09nanoSQUIDs}.
This approach typically produced SQUIDs with $S_\Phi^{1/2}\,\gapprox\,1\,{\rm m}\Phi_0/\sqrt{\rm Hz}$.
Only very recently, highly sensitive Nb SQUIDs with $S_\Phi^{1/2}\approx 0.2\mu\Phi_0/\sqrt{\rm Hz}$ have been demonstrated \cite{Hao08}.
%
These encouraging results have been obtained with FIB patterned constriction junctions at $T=6.8\,$K.

For operation in high magnetic fields $B$ a small junction size ($\perp \vec{B}$) is required due to the suppression of $I_c$ with increasing magnetic flux in the junction above several $\Phi_0$, which demands for a particularly large $j_c$.
Furthermore, the maximum field of operation is limited by the upper critical field $B_{c2}$, e.~g.~to $\lapprox\,1\,$T for typical Nb thin film based SQUIDs.
Here, high-transition-temperature (high-$T_c$) SQUIDs offer three advantages: (i) very high $B_{c2}$ in the tens of Tesla regime or even more, (ii) high $j_c > 10^5\,{\rm A/cm}^2$ for grain boundary junctions (GBJs)
operating at $T=4.2\,$K and well below and (iii) a GBJ geometry with the junction barrier perpendicular to the thin film SQUID loop; this allows to apply very large in-plane fields (for switching the magnetization of nanoparticles) which do not couple to the SQUID and which do not reduce $I_c$.
Here, the challenge is to produce sub-$\mu$m GBJs with high quality, in particular with high $j_c$ and $\rho=RA_J$, where $R$ is the junction resistance.
%
At this point, we should mention, that in principle, also constriction type junctions based on HTS, or MgB$_2$ thin films may fulfill the above mentioned requirements, and in fact such junctions and SQUIDs based on them have been patterned by FIB \cite{Blank95,Pedyash96,Brinkman01,Burnel02a,Wu08b}. However, these junctions typically often show flux-flow type or hysteretic IVCs at low $T$, such that their operation temperature is often limited to a narrow $T$ range well above 4\,K, and their performance with respect to flux noise so far did never reach the performance of high-$T_c$ low-noise GBJ SQUIDs \cite{Koelle99}.

Already in the 1990s thin film high-$T_c$ YBa$_2$Cu$_3$O$_7$ (YBCO) sub-$\mu$m GBJs and SQUIDs have been fabricated using e-beam lithography \cite{Kawasaki91,Elsner98}.
However, oxygen loss during processing, in particular for very thin films, may require post-deposition annealing to improve junction characteristics \cite{Herbstritt01}.
More recently, sub-$\mu$m YBCO GBJs have also been fabricated by FIB \cite{Testa04}, and both technologies enabled fundamental studies on transport and noise in high-$T_c$ sub-$\mu$m GBJs \cite{Herbstritt01,Ilichev01,Tzalenchuk03,Testa05}.
Still, a significant degradation of $j_c$ for $w\,\lapprox\,500\,$nm was found
\cite{Kawasaki91,Elsner98,Testa04}, and the use of deep sub-$\mu$m GBJs for the realization of nanoSQUIDs has not been explored yet.
%
The motivation for the realization of sub-$\mu$m GBJs with widths well below 500\,nm is based on the following considerations: First of all, operation in high magnetic fields in the Tesla range as mentioned above, requires very accurate alignment of the applied magnetic field in the thin film plane in order to reduce coupling of the applied out-of-field component to the GBJ. This demands for as small as possible GBJ widths. Furthermore, as will be shown below, the spin sensitivity scales linearly with the rms flux noise $S_\Phi^{1/2}$ of the SQUID. Optimization of $S_\Phi^{1/2}$ requires an as small as possible SQUID loop inductance $L$ \cite{Tesche77}, i.e.~minimization of the dimensions of the SQUID loop, which for topological reasons has to be intersected by the grain boundary. Here, SQUID loop sizes of the order of 100\,nm seem to be feasible, according to our experience on FIB patterning of our devices as described below. In order to ensure optimum SQUID performance, one should achieve at least a few SQUID modulations within the Fraunhofer-like $I_c(B)$ modulation of the single GBJs. This in turn requires also shrinking the GBJ widths down to the size of the SQUID loop. Hence, our goal is to demonstrate the feasibility of FIB patterning YBCO GBJs down to junction widths of the order of 100\,nm. In order to accomplish this, we investigated the scaling behavior of YBCO GBJ properties with linewidths ranging over two orders of magnitude, from $\sim 8\,\mu$m down to 80\,nm, and we investigated the electric transport and noise properties of YBCO GBJ dc SQUIDs with the smallest linewidths achieved within this study.
%
We note that we performed so far only investigations on low-field properties of the fabricated GBJs and SQUIDs, as we are at this stage interested in clarifying the intrinsic scaling properties of our devices with GBJ width, although the ultimate goal of this work is to
operate such SQUIDs in high magnetic fields in the Tesla range at $T=4.2\,K$ and well below.

%
The remainder of this paper is organized as follows.
Section 2 
very briefly addresses sample fabrication and layout, including some information on the quality of our YBCO thin films.
In Sect.~3 
we first describe and discuss the results of electric transport properties of our shunted GBJs, with focus on their dependence on junction width, which was varied over two orders of magnitude (3.1). 
The second part (3.2) 
of this section describes the results obtained for our SQUIDs, with focus on electric transport and noise properties of the SQUID with the smallest GBJ width used in this study (SQUID 2).
Having characterized our SQUIDs, we discuss in Sect.~4 
the important relation between the flux noise $S_\Phi^{1/2}$ of the SQUIDs and the spin sensitivity $S_\mu^{1/2}$, which is the important figure of merit for detection of small spin particles.
Here, we provide a solution for calculating the spin sensitivity for any arbitrary geometry of the SQUID loop as a function of position and orientation of the magnetic moment of a small particle to be detected.
We then apply this solution to the particular geometry of SQUID 2 and finally discuss perspectives for further optimization of $S_\mu^{1/2}$.
Section 5 contains our conclusions.

\section{Sample Fabrication}

\begin{figure}[b]
\ifgraph\includegraphics[width=15cm]{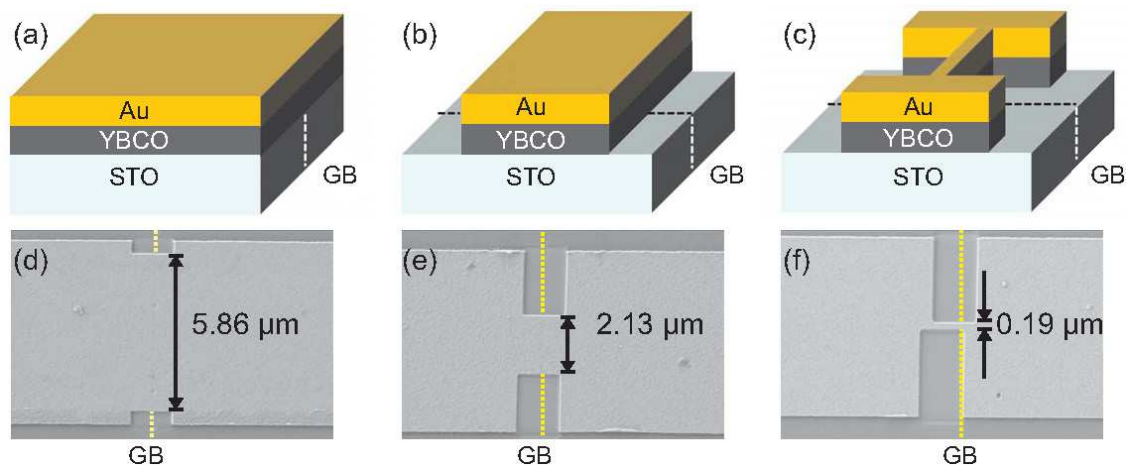}\fi
\caption{(Color online) Upper row: Schematic illustration of the steps used for fabricating YBCO grain boundary junctions (GBJs). (a) in-situ deposition of a YBCO/Au bilayer on a bicrystal STO substrate; (b) patterning of $8\,\mu$m wide bridges straddling the grain boundary (GB) (by photolithography and Ar ion milling); (c) patterning of a narrow GBJ by FIB.
The location of the GB is indicated by dashed lines.
Bottom row (d)--(f) shows scanning electron microscopy images of three single junction devices (from chip 1) with different width (indicated in the graphs).}
\label{Fig:1}
\end{figure}

\begin{figure}[tb]
\ifgraph\includegraphics[width=15cm]{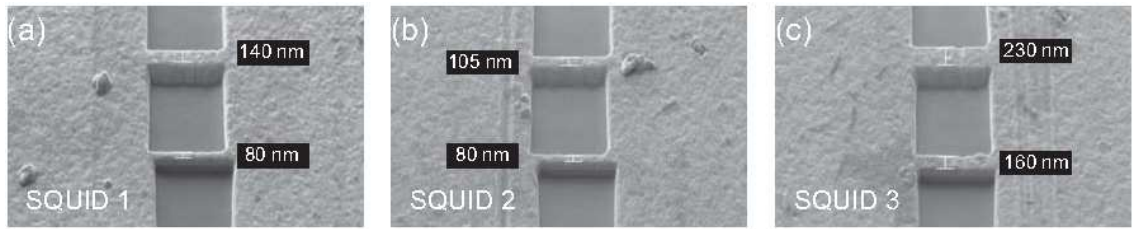}\fi
\caption{SEM images of the three SQUIDs (loop size $1.0\times 1.2\,\mu{\rm m}^2$) fabricated on chip 2.
Labels in black boxes give junction widths.}
\label{Fig:2}
\end{figure}

%
We fabricated devices on SrTiO$_3$ (STO) symmetric [001] tilt bicrystal substrates with misorientation angle $\theta=30\,^\circ$ (chip 1) and $24\,^\circ$ (chip 2).
Figure \ref{Fig:1}(a)--(c) illustrates the fabrication steps.
We deposited $d_Y=50\,$nm thick $c$-axis oriented epitaxially grown YBCO by pulsed laser deposition (PLD), followed by in-situ evaporation of Au with thickness $d_{Au}=20\,$nm (chip 1) and 60\,nm (chip 2), serving as a resistive shunt and protection layer during FIB milling.
For details on PLD growth of our YBCO films on STO substrates, and their structural and electric transport properties see Ref.~\cite{Werner-YBCO-LCMO}.
In brief, our 50\,nm thick YBCO films typically yield $0.1^\circ$ full width half maximum of the rocking curve at the (005) x-ray diffraction peak, have $T_c=91\,$K with a transition width $\sim 0.5\,$K and resistivity $\rho\approx 50\,\mu\Omega$cm at $T=100\,$K.
On both chips,
$8\,\mu$m wide bridges straddling the grain boundary were fabricated by photolithography and Ar ion milling and then patterned by FIB with Ga ions (50\,pA, 30\,kV) to make junctions and dc SQUIDs with junction widths $80\,{\rm nm}\le w\le 7.8\,\mu$m.
%
%
FIB patterning was performed with a dual beam 1540 XB cross beam (Zeiss).
This allowed us to apply an optimized FIB cut procedure (soft FIB procedure), with small ion current density and minimum ion exposure time of non-milled areas, i.e.~only very brief snap shot imaging prior to milling.
Even for imaging by the electron beam, we minimized the exposure time in order to avoid damaging of our FIB cut bridges.
Figure \ref{Fig:1}(d)--(f) shows scanning electron microscopy (SEM) images of three GBJs fabricated on chip 1.
In total, we investigated 22 single GBJs and 3 dc SQUIDs (on chip2; hole size  $1.0\times 1.2\,\mu{\rm m}^2$).
SEM images of the three SQUIDs are shown in Fig.~\ref{Fig:2}.

\section{Experiments}


We characterized our devices at $T=4.2\,$K in a magnetically shielded environment.
For measurements of IVCs, $I_c(B)$ and $V(B)$ we used a 4-point arrangement with a room temperature voltage amplifier.
For SQUID noise measurements we preamplified the output signal with a Nb dc SQUID amplifier with $ 0.1\,{\rm nV}/\sqrt{\rm Hz}$ resolution.

\subsection{Transport properties vs junction width}

\begin{figure}[tb]
\ifgraph\includegraphics[width=7cm]{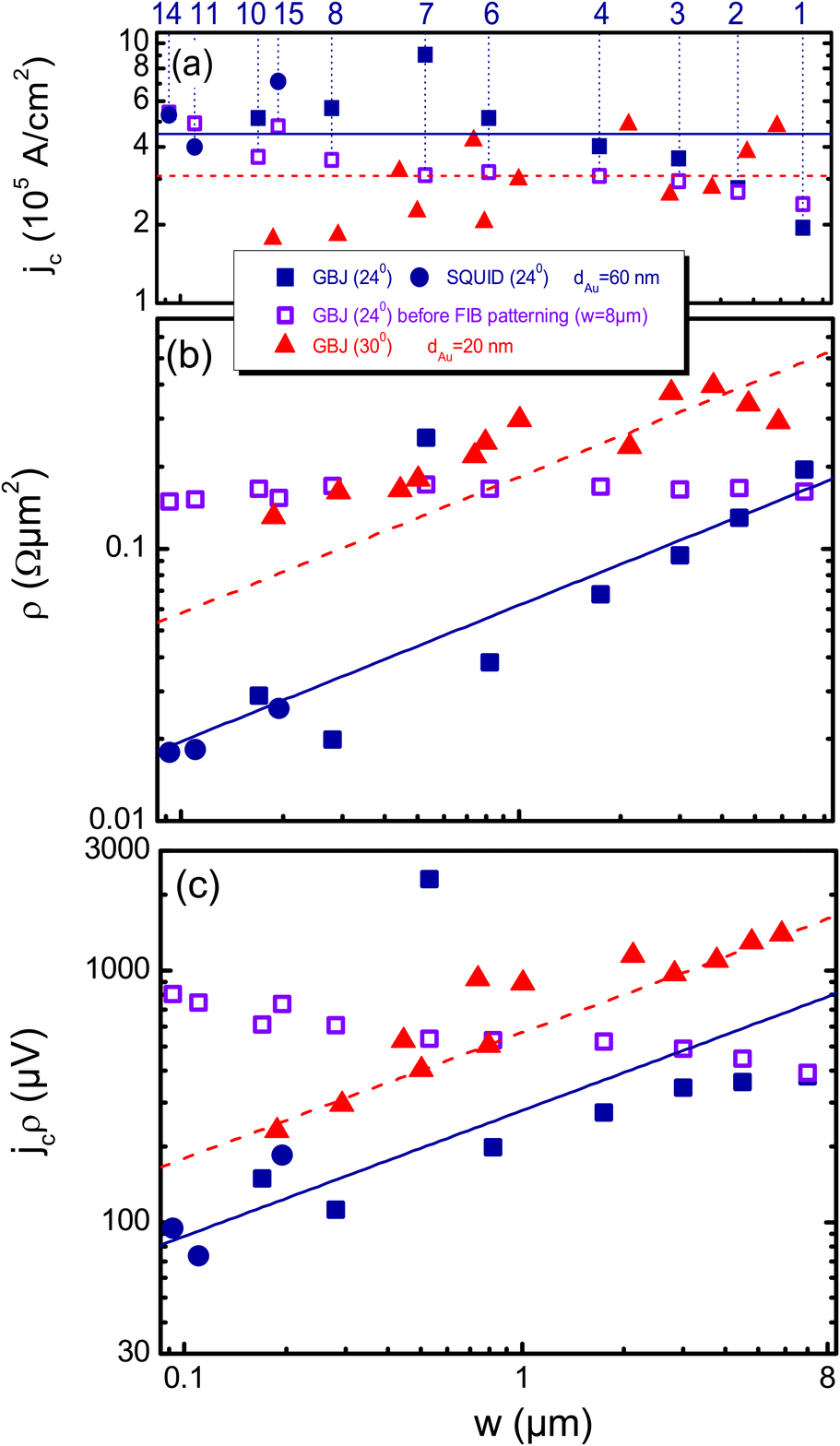}\fi
\caption{(Color online) Transport data of YBCO GBJs and SQUIDs vs junction width $w$ (solid symbols):
(a) critical current density $j_c(w)$;
(b) junction resistance times area $\rho(w)$;
(c) $j_c\rho(w)$.
Dashed (solid) lines indicate average $j_c$ values [in (a)] and approximate scaling of $\rho(w)$ [in (b)] and $j_c\rho(w)$ [in (c)] for chip 1 (2).
Open squares are data for the same GBJs on chip 2, measured prior to FIB patterning (i.e. $w=8\,\mu$m).
The numbers on the top axis in (a) label the device numbers on chip 2.}
\label{Fig:xport}
\end{figure}

All devices showed resistively-and-capacitively-shunted-junction (RCSJ)-type IVCs, which for some of the sub-$\mu$m junctions on chip 1 (thinner Au shunt) had a small hysteresis.
Therefore, for chip 2, we increased $d_{Au}$ by a factor of three, yielding non-hysteretic IVCs, except for the 530\,nm wide junction, which has an exceptionally high $j_c\rho$.
From the IVCs we determined $I_c$, and $R$, and calculated $j_c$ and $\rho$, using $d_Y=50\,$nm and $w$ as obtained from SEM images.
The results of these measurements are summarized in Fig.~\ref{Fig:xport}, plotted vs $w$ which spans two orders of magnitude.
Full symbols in Fig.~\ref{Fig:xport} show data obtained after FIB patterning; open symbols show data (for chip 2) from the $8\,\mu$m wide bridges prior to FIB patterning.

Figure \ref{Fig:xport}(a) shows $j_c(w)$, which is well above $10^5\,{\rm A/cm^2}$ over the entire range of $w$.
The $j_c$ values shown are typical for $\theta=24^\circ$ and $30^\circ$ YBCO GBJs at 4.2\,K and $w\,\gapprox\,2\,\mu$m\cite{Koelle99}; however, such high $j_c$ values have not been previously observed for widths down to 80\,nm.
We do find a significant scattering of $j_c(w)$, however without a clear width dependence.
The average $j_c$ for chip 2 is 1.5 times the one for chip 1, as expected from the scaling $j_c(\theta)$ of GBJs \cite{Hilgenkamp02}.
The comparison of $j_c$ of the same bridges before and after FIB patterning shows that for most devices $j_c$ even slightly increased after FIB patterning.
The position of the devices on chip 2 (along the GB of the substrate) is ordered according to their device number (1--15) [c.f.~top axis of Fig.~\ref{Fig:xport}(a)] from the left to the right edge of the substrate.
There is a clear trend of increasing $j_c$ by about a factor of two (for the $8\,\mu$m wide bridges) along the entire substrate.
The origin of this gradient in $j_c$ has not been clarified; however we can rule out a corresponding variation in the YBCO film thickness across the substrate.
%
A possible explanation for the observed gradient in $j_c$ of the $8\,\mu$m wide GBJs along chip 2, could be a gradient in the quality of the GB in the bicrystal substrate, which in turn can cause a gradient in the barrier thickness of the GBJs along the chip.

Figure \ref{Fig:xport}(b) shows an approximate scaling $\rho\propto\sqrt{w}$ of unclear origin.
%
We note that the lines shown in Fig.~\ref{Fig:xport}(b) and Fig.~\ref{Fig:xport}(c) are no fits to our data. These lines are just drawn to illustrate the trend of decreasing resistivity $\rho$ with decreasing junction with $w$.
Before FIB patterning, the  $8\,\mu$m wide GBJs on chip 2 all had $\rho\approx 0.17\,\Omega\mu{\rm m}^2$, which falls onto the observed $\rho(w)$ dependence, indicating that this scaling is not specific to FIB patterned GBJs.
Furthermore, $\rho\approx 0.17\,\Omega\mu{\rm m}^2$ is an order of magnitude below typical values for unshunted GBJs \cite{Hilgenkamp02}, which we attribute to the Au shunt, and which is also consistent with the larger $\rho$ of GBJs on chip 1 with thinner Au.
For chip 1, $d_{Au}=20\,$nm is close to the 15\,nm implantation depth of 30\,keV Ga ions in Au \cite{Rubanov03}.
Hence one might expect that FIB induces an increase in the Au resistivity via Ga implantation.
This effect should be suppressed for chip 2 with 3 times thicker Au.
In any case, it is hard to explain, why Ga implantation should increase $\rho$ for wider junctions.
%
Certainly, Ga implantation is not the only detrimental effect of FIB patterning. In particular, the Ga beam might destroy the crystalline order close to the patterned edges. However, our experimental observation of almost constant $j_c$ for GBJ widths down to 80\,nm rules out severe edge damage effects on a length scale of several tens of nm. This observation also rules out such effects as a possible explanation for the observed scaling behavior of $\rho(w)$.
%

Figure \ref{Fig:xport}(c) shows $j_c\rho(w)\approx 0.1\ldots 1\,$mV, i.e.~at least one order of magnitude below the values for unshunted YBCO GBJs of comparable $j_c$. This is certainly due to the suppression of $j_c\rho$ by the Au shunt required to ensure non-hysteretic IVCs at 4.2\,K.
The decrease of $j_c\rho$ with decreasing $w$ is due to the scaling of $\rho(w)$ mentioned above.
Still, even for the 80\,nm wide GBJs we find reasonable values of $j_c\rho$ around $100\,\mu$V, which are certainly quite suitable for the realization of sensitive SQUIDs.

\subsection{SQUID parameters, transport characteristics and noise performance}

\begin{table}[b]
\begin{tabular}{ccccccccc}
\hline\hline
\rule{0mm}{4mm}
\#  & $w_1$ & $w_2$ & $I_c$     & $R$        & $I_cR$   & $L$  & $\beta_L$ & $V_\Phi$          \\ 
    & [nm]  & [nm]  & [$\mu$A]  & [$\Omega$] & [$\mu$V] & [pH] &           & [mV$/\Phi_0$] \\ 
\hline
1   & 80    & 140   & 44        &  1.7       & 73       & 15   & 0.31      & 0.08              \\ 
2   & 80    & 105   & 49        &  1.9       & 94       & 16   & 0.38      & 0.11              \\ 
3   & 160   & 230   & 139       &  1.3       & 185      & 10   & 0.66      & 0.13              \\ 
\hline\hline
\end{tabular}
\caption{Parameters of YBCO GBJ dc SQUIDs.}
\label{tab:SQUIDs}
\end{table}

The results of transport measurements on all three SQUIDs (on chip 2) are summarized in Tab.~\ref{tab:SQUIDs}.
The SQUID inductance $L$ was calculated with the numerical simulation software 3D-MLSI \cite{Khapaev03}, which is based on a finite element method to solve the London equations for a given film thickness and London penetration depth ($d_Y=50\,$nm and $\lambda_L=140\,$nm, respectively, in our case).
As $d_Y\ll\lambda_L$, the kinetic inductance contributes significantly to $L$.
For SQUID 1 and SQUID 2, the GBJ widths $w_i$ ($i=1,2$) are below $\lambda_L$, which increases $L$ over the one of SQUID 3 with wider junctions.
From the calculated $L$ and measured $I_c$ we obtain $\beta_L\equiv LI_c/\Phi_0\approx0.3\ldots 0.7$, i.~e.~not far from the optimum value $\beta_L\approx 1$ \cite{Kleiner04}.
The transfer function $V_\Phi$, i.~e.~the slope of the $V(\Phi)$ curves at optimum bias current and applied flux $\Phi=\pm\frac{1}{4}\Phi_0$ is around $0.1\,{\rm mV}/\Phi_0$, and the effective area $A_{eff}=\Phi/B\approx 8\,\mu$m$^2$ for all three SQUIDs.

\begin{figure}[t]
\ifgraph\includegraphics[width=15cm]{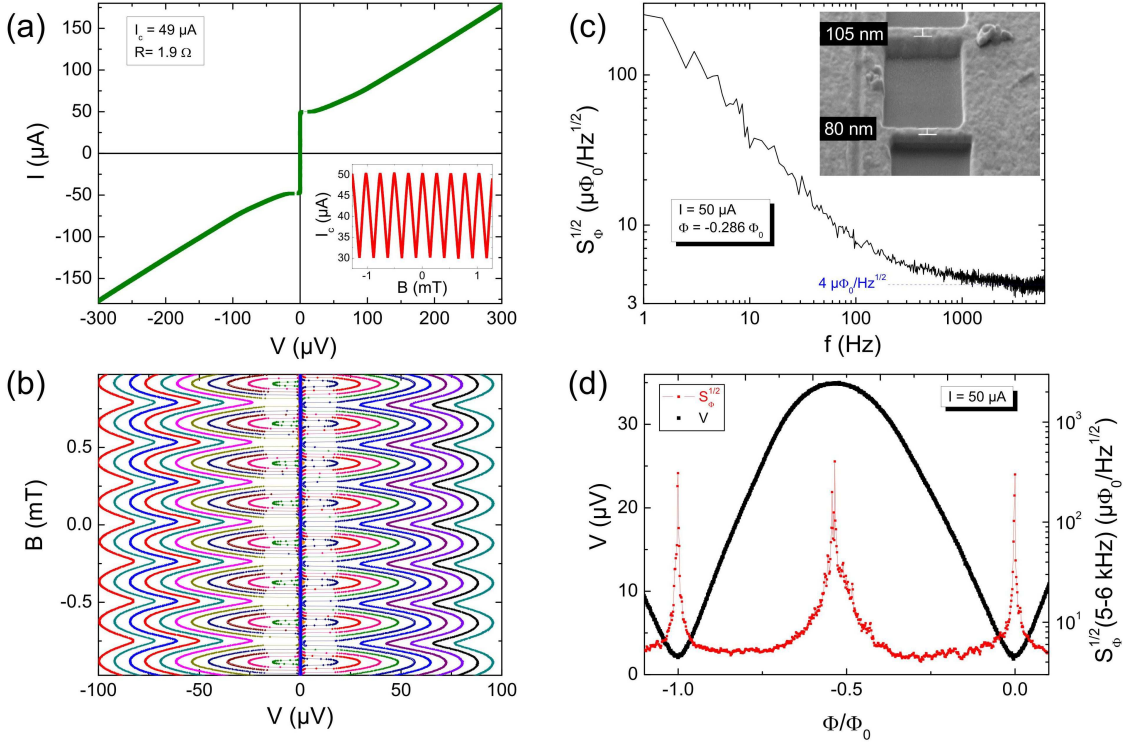}\fi
\caption{(Color online) Transport and noise characteristics of SQUID 2:
(a) IVC at $B=0$; inset shows $I_c(B)$.
(b) $V(B)$ for $I=-70.8\ldots-12.3$ and $10.0\ldots 68.5\,\mu$A (in $3.9\,\mu$A steps).
(c) Spectral density of rms flux noise $S_{\Phi}^{1/2}(f)$; inset: SEM image of the SQUID.
(d) $V(\Phi)$ and $S^{1/2}_{\Phi}(\Phi)$ (averaged from $f=$ 5 to 6\,kHz).}
\label{Fig:SQUID2}
\end{figure}

Figure \ref{Fig:SQUID2} shows electric transport and noise data obtained for SQUID 2 [the device with smallest $w$; see inset in (c)].
Figure \ref{Fig:SQUID2}(a) shows an IVC for an applied field $B=0$ corresponding to a maximum in $I_c$.
The small jump at $I_c$ to $V\neq 0$ indicates that the junctions are at the transition to the underdamped regime.
The inset in Fig.~\ref{Fig:SQUID2}(a) shows $I_c(B)$ with 40\,\% modulation.
Figure \ref{Fig:SQUID2}(b) shows $V(B)$ for various bias currents $I$.
The small shift in the minima of $V(B)$ upon reversing $I$ is in accordance with the $I_c$ asymmetry of the two GBJs due to their different widths.

Finally, graphs (c) and (d) in Fig.~\ref{Fig:SQUID2} show the results of noise measurements on SQUID 2.
Fig.~\ref{Fig:SQUID2}(c) shows the rms spectral density of flux noise $S_\Phi^{1/2}(f)\propto f^{-x}$ for optimum flux bias $\Phi=-0.286\,\Phi_0$ with $x\approx 1.8$ for frequencies $f\,\lapprox\,1\,$kHz, which we attribute to $I_c$ fluctuations in the GBJs \cite{Koelle99}.
For larger $f$ we find a white flux noise level $S_{\Phi,w}^{1/2}\approx 4\,\mu\Phi_0/\sqrt{\rm Hz}$, which to our knowledge is the lowest value of $S_\Phi^{1/2}$ obtained for a YBCO dc SQUID with sub-$\mu$m GBJs so far.
Fig.~\ref{Fig:SQUID2}(d) shows the rms flux noise $S_{\Phi}^{1/2}$ (averaged from $f=$ 5 to 6\,kHz) and the SQUID voltage $V$ vs applied flux $\Phi$.
We find a rather shallow minimum in $S_{\Phi}^{1/2}(\Phi)$ for an applied flux where the slope of the $V(\Phi)$ curve [also shown in graph (d)] is close to its maximum.


\section{Spin sensitivity}

Coming back to the main motivation of this work, i.e.~the development of nanoSQUIDs for the detection of small spin systems, we derive an expression for the spin sensitivity $S_\mu^{1/2}$, which we then use to calculate $S_\mu^{1/2}$ for the particular geometry and flux noise of SQUID 2 as a function of the position of a magnetic particle for a given orientation of its magnetic moment.
$S_\mu$ is the spectral density of spin noise, which depends on the spectral density of flux noise $S_\Phi$ of the SQUID and on the coupling between a magnetic particle with magnetic moment $\vec{\mu}=\mu\cdot\hat{e}_\mu$ and the SQUID via the relation $S_\mu=S_\Phi/\phi_\mu^2$.
Here, $\phi_\mu(\hat{e}_\mu,\vec{r}_\mu)\equiv \Phi_\mu(\vec{\mu},\vec{r}_\mu)/\mu$ is the magnetic flux $\Phi_\mu$ per magnetic moment $\mu$ coupled into the SQUID loop by the magnetic particle, which is located at the position $\vec{r}_\mu$ and which is oriented along $\hat{e}_\mu$.
This means, in order to determine $S_\mu^{1/2}$ for a given $S_\Phi^{1/2}$, one needs to calculate the coupling function $\phi_\mu(\hat{e}_\mu,\vec{r}_\mu)$, which will also depend on the SQUID geometry.

To determine  $\phi_\mu$, we assume that the magnetic moment $\vec{\mu}$ is moved from a distance far away to a position $\vec{r}=\vec{r}_\mu$ close to the SQUID loop.
When the magnetic moment approaches $\vec{r}_\mu$, a circulating current $I_\mu(\vec{\mu},\vec{r}_\mu)$ is induced in the SQUID loop, which compensates the coupled flux $\Phi_\mu$, due to the diamagnetic response of the SQUID loop.
The magnetic field energy stored in the loop of inductance $L$ is $W_{loop}=\frac{1}{2}LI_\mu^2$.
The work required to place the particle in the magnetic field $\vec{B}_\mu(\vec{r})$ produced by the circulating current $I_\mu$ is $W_\mu=-\frac{1}{2}\vec{\mu}\cdot\vec{B}_\mu(I_\mu,\vec{r}_\mu)$.
We note that $W_\mu>0$, due to the diamagnetic response of the SQUID loop.
Hence, the total work required to bring the magnetic particle to the position $\vec{r}_\mu$ is
\begin{equation}
W_1=W_{loop}+W_\mu=\frac{1}{2}LI_\mu^2 -\frac{1}{2} \vec{\mu}\cdot\vec{B}_\mu(I_\mu,\vec{r}_\mu)\;.
\label{eq:W1}
\end{equation}
%
On the other hand, instead of the SQUID, we may consider a fixed current system producing the same field $\vec{B}_\mu(I_\mu,\vec{r}_\mu)$ as the SQUID, when the particle is in its final position $r_\mu$.
In this case, the particle has a (positive) energy
\begin{equation}
W_2=-\vec{\mu}\cdot\vec{B}_\mu(I_\mu,\vec{r}_\mu)\;.
\label{eq:W2}
\end{equation}
%
From $W_1=W_2$ we obtain $I_\mu^2=-\vec{\mu}\cdot\vec{B}_\mu/L$.
With $\Phi_\mu=LI_\mu$ and with $\vec{B}_\mu/I_\mu=\vec{B}/I\equiv \vec{b}$ one thus obtains
\begin{equation}
\frac{\Phi_\mu(\vec{\mu},\vec{r}_\mu)}{\mu}\equiv \phi_\mu(\hat{e}_\mu,\vec{r}_\mu) = -\hat{e}_\mu\cdot\vec{b}(\vec{r}_\mu)\;,
\label{eq:coupling}
\end{equation}
where $I$ is an arbitrary current circulating in the SQUID loop, which generates the magnetic field $\vec{B}(I)$ at the position $\vec{r}_\mu$ of the magnetic particle.

\begin{figure}[t]
\ifgraph\includegraphics[width=15cm]{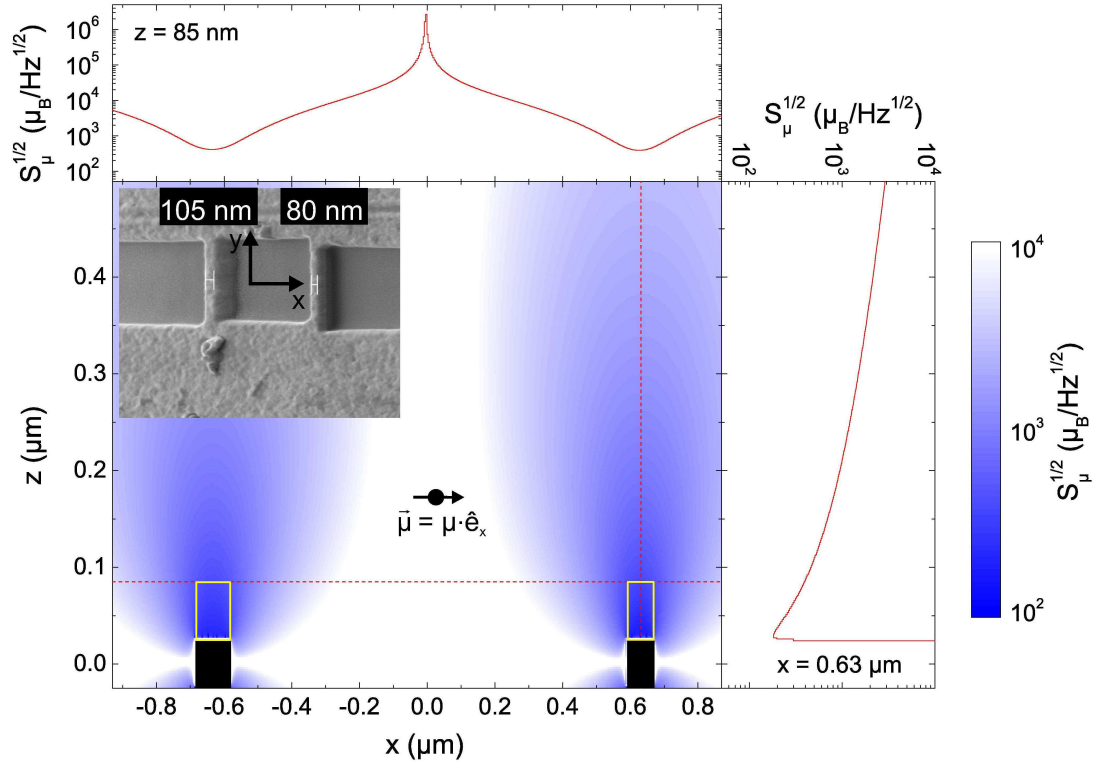}\fi
\caption{(Color online) Calculated spin sensitivity $S^{1/2}_{\mu}$ for SQUID 2 with $S^{1/2}_{\Phi}=4\,\mu\Phi_0/\sqrt{\rm Hz}$ for the detection of the magnetic moment of a small spin particle aligned along the $x$ axis in the $(x,y)$ plane of the SQUID loop [c.f.~inset in main graph].
Main graph shows a contour plot of $S^{1/2}_{\mu}$ as a function of the position of the particle in the $(x,z)$ plane at $y=0$.
The location of the YBCO bridges is indicated by the black rectangles; the rectangles (yellow lines) on top of those indicate the position of the Au layer.
The linescan above the main graph shows $S^{1/2}_{\mu}(x)$ for $z=85\,$nm, and the linescan right to the main graph shows $S^{1/2}_{\mu}$(z) for $x=0.63\,\mu$m.
The location of these line scans is indicated by the dashed (red) lines in the main graph.}
\label{Fig:Spinsensitivity}
\end{figure}

Equation (\ref{eq:coupling}) reproduces the results of Refs.~\cite{Ketchen89,Tilbrook09}, derived for a circular filamentary SQUID loop.
Moreover, Eq.~(\ref{eq:coupling}) provides a solution of the problem, valid for any \textit{arbitrary} geometry of the superconducting loop, if one can find the normalized magnetic field distribution $\vec{b}(\vec{r})$ outside the SQUID loop.
For a given $\vec{b}(\vec{r})$ (determined by the SQUID geometry only) and given flux noise, one can use Eq.~(\ref{eq:coupling}) to easily calculate the spin sensitivity $S_\mu^{1/2}=S_{\Phi}^{1/2}/\phi_\mu$ for any orientation $\hat{e}_\mu$ and location of the magnetic particle.

For the geometry of SQUID 2, we calculated the spatial distribution of the current density in the SQUID loop and the corresponding 3-dimensional magnetic field distribution $\vec{b}(\vec{r})$ outside the SQUID loop with 3D-MLSI \cite{Khapaev03}.
Figure \ref{Fig:Spinsensitivity} shows the resulting spin sensitivity of
SQUID 2 (with $S^{1/2}_{\Phi}=4\,\mu\Phi_0/\sqrt{\rm Hz}$) for the detection of a magnetic particle located in the $(x,z)$ plane (at $y=0$) with its magnetic moment pointing along the $x$ direction, i.e.~$\hat{e}_\mu=\hat{e}_x$.
I.e., the magnetic moment of the particle is aligned parallel to the thin film plane of the SQUID, and perpendicular to the current through the GBJs.

The contour plot of the spin sensitivity shows clear minima right above
the superconducting bridges straddling the grain boundary.
The upper inset shows a linescan $S_\mu^{1/2}(x)$ of the spin sensitivity
at a height $z=d_{Au}+d_Y/2=85\,$nm above the ring (i.e. for our SQUID the
minimum vertical distance due to the Au layer on top of the YBCO film).
The lowest value of the spin sensitivity along this linescan is $390\,\mu_B/\sqrt{\rm Hz}$, which could be further improved by reducing the thickness of the Au layer.
This can be done even without affecting the GBJ properties if the Au layer is not removed right above the GBJ.
Removing the gold layer (and placing the magnetic particle at $z=d_Y/2=25\,$nm) would improve $S_\mu^{1/2}$ by more than a factor of two down to $180\,\mu_B/\sqrt{\rm Hz}$,
as can be seen in the right inset, which shows the vertical dependence $S_\mu^{1/2}(z)$ at $x=0.63\,\mu$m, i.e.~right above the center of the YBCO bridge.
Moreover, further improvements in $S_\mu^{1/2}$ are feasible by improving $S_\Phi^{1/2}$, which is by no means optimized for the SQUIDs presented here.
%
%
For example, our FIB technology allows for a reduction in the size of the SQUID loop down to $\sim 100\,$nm and a concomitant reduction in SQUID inductance $L$ down to $\sim 1\,$pH.
This, in turn, can lead to a significant improvement in $S_\Phi^{1/2}$ by at least an order of magnitude, which would bring $S_\mu^{1/2}$ down to $\approx 20\,\mu_B/\sqrt{\rm Hz}$.

\section{Conclusions}

In conclusion, we have fabricated YBCO grain boundary junctions and dc SQUIDs by FIB patterning with junction widths ranging from $7.8\,\mu$m down to 80\,nm.
Using an Au thin film shunt on top of the junctions, we achieved non-hysteretic current-voltage-characteristics for operation of YBCO dc SQUIDs at 4.2\,K
and below.
We demonstrated that FIB pattering enables the fabrication of deep sub-$\mu$m GBJs without degradation of critical current densities, and comparable to GBJs with widths above $1\,\mu$m.
We do find a systematic dependence of the resistance times area $\rho$ of our GBJs, which scales approximately with the junction width $w$ as $\rho\propto\sqrt{w}$.
The origin of this scaling could not be resolved and requires further studies.
Still, we obtain values of $I_cR$ for our GBJs around $100\,\mu$V for junctions on the 100\,nm scale, which is promising for the fabrication of sensitive nanoSQUIDs.
We demonstrated low-noise performance for such devices in the $\mu\Phi_0/\sqrt{\rm Hz}$ range, which still can be improved significantly and which makes them promising candidates for applications in magnetic nanoparticle detection and measurements at high magnetic fields.
%
%
The presented solution for calculating the spin sensitivity for arbitrary SQUID geometries -- as a function of position and orientation of the magnetization of small spin particles -- provides an important tool for the systematic optimization of the spin sensitivity using nanoSQUIDs as sensitive devices for direct detection of magnetization switching of small spin particles.

\ack

We gratefully acknowledge Thomas Dahm for helpful discussions.
J.~Nagel and M.~Kemmler acknowledge support by the Carl-Zeiss Stiftung, K.~Konovalenko acknowledges support by the Otto Benecke Stiftung and R.~Werner acknowledges support by the Cusanuswerk, Bisch\"ofliche Studienf\"orderung.
This work was funded by the Deutsche Forschungsgemeinschaft (DFG) via the SFB/TRR 21.

\section*{References}

\bibliographystyle{unsrt}

\bibliography{Nagel-FIB-GBJ-SQUID}

\end{document}